\documentclass[]{spie}

\usepackage{amsmath,amsfonts,amssymb}
\usepackage{graphicx}
\usepackage[colorlinks=true, allcolors=blue]{hyperref}
\usepackage{subcaption}
\usepackage{todonotes}
\usepackage{textcomp}

\graphicspath{{res/}}

\title{Automatic Online Quality Control of Synthetic CTs}

\author[a,b,d]{Louis D. van Harten}
\author[a,d]{Jelmer M. Wolterink}
\author[b]{Joost J.C. Verhoeff}
\author[a,c,d]{Ivana I\v{s}gum}

\affil[a]{Department of Biomedical Engineering and Physics, Amsterdam University Medical Center, University of Amsterdam, The Netherlands}
\affil[b]{Department of Radiotherapy, University Medical Center Utrecht, Utrecht, The Netherlands}
\affil[c]{Department of Radiology and Nuclear Medicine, Amsterdam University Medical Center, University of Amsterdam, The Netherlands}
\affil[d]{Image Sciences Institute, University Medical Center Utrecht, Utrecht, The Netherlands}

\authorinfo{Further author information: (Please send correspondence to L.D. van Harten)\\E-mail: l.d.vanharten@amsterdamumc.nl, Telephone: +31 20 56 65206}

\begin{document}

\maketitle

\begin{abstract} 
    Accurate MR-to-CT synthesis is a requirement for MR-only workflows in radiotherapy (RT) treatment planning. In recent years, deep learning-based approaches have shown impressive results in this field. However, to prevent downstream errors in RT treatment planning, it is important that deep learning models are only applied to data for which they are trained and that generated synthetic CT (sCT) images do not contain severe errors. For this, a mechanism for online quality control should be in place.
    In this work, we use an ensemble of sCT generators and assess their disagreement as a measure of uncertainty of the results. We show that this uncertainty measure can be used for two kinds of online quality control. First, to detect input images that are outside the expected distribution of MR images. Second, to identify sCT images that were generated from suitable MR images but potentially contain errors. Such automatic online quality control for sCT generation is likely to become an integral part of MR-only RT workflows.
\end{abstract}


\section{Introduction}
In recent years, substantial work has been published exploring the possibilities for MR-only workflows in radiotherapy (RT) treatment\cite{edmund2017review}. 
Radiotherapy (RT) treatment planning typically requires patients to undergo an MR scan for soft tissue imaging and a CT scan to estimate radioactive dose absorption of locations in and around the treatment target volumes. To reduce costs and ionizing radiation for the patient, as well as to speed up the RT treatment planning process, it would be desirable to omit CT acquisition from the treatment workflow. The recent availability of MR-Linac machines has fueled the interest in such MR-only RT treatment\cite{lagendijk2014magnetic}. An MR-only workflow replaces the CT image by a synthetic CT (sCT) image generated from the MR image. Impressive results for sCT generation have been achieved using deep learning, in particular using generative adversarial networks (GANs) \cite{edmund2017review, wolterink_deep_2017, nie2017medical}.

For MR-only workflows to be adopted in the clinic, CT synthesis should be accurate and robust. An sCT generator is a model which defines a mapping between the data distribution of MR images and the data distribution of CT images, but there is no exact analytical form of this mapping. Deep learning-based methods for sCT generation perform well when the input at test time is from the same distribution as the training data, but they may fail when applied to an MR image that falls outside this distribution. This could be the case when severe pathology is present or when patients are scanned with a different MRI machine due to logistics or maintenance. More robust sCT generative models could be obtained by using larger and more diverse training sets, i.e. by providing a wider data distribution at training time. However, it cannot be guaranteed that the entire space of possible inputs is covered. In case an input falls outside of the training distribution, an online quality control mechanism should be in place to alert a human operator, minimizing the chances of severe errors in the automatically generated synthetic CT images causing inaccurate treatment plans\cite{nyholm2014counterpoint}.

In this work, we propose a method for such quality control in 2D CNN-based sCT synthesis models by evaluating the model uncertainty. Uncertainty in sCT generation has previously been explored in the context of Gaussian mixture regression \cite{johansson2012voxel} and deep learning with Bayesian sampling \cite{bragman2018uncertainty}. Here, we determine uncertainty as the voxel-wise ensemble disagreement among three GANs trained for sCT generation of the head. Such a method for uncertainty estimation was originally proposed as an efficient alternative to Bayesian sampling, showing promising results on natural imaging tasks\cite{lakshminarayanan2017simple}. We build on this method by training the three different sCT generators in the ensemble on different views of the data (i.e. axial, coronal or sagittal slices). This means each network is trained to estimate a mapping between a different set of data distributions. This prevents the three networks from learning an identical function, which would result in all networks making identical mistakes when presented with out-of-distribution input data.

We train an ensemble of three GANs using T1-weighted images and evaluate whether our uncertainty metric can differentiate between three kinds of inputs at test time: T1-weighted images from the same distribution as the training set, T1-weighted images with  gadolinium-based contrast agent acquired on the same scanner, and T1-weighted images acquired using a different sequence on a different model scanner. We show that based on the uncertainty of the model, we can identify unsupported inputs, such as MR images from a different scanner, that may lead to erroneous outputs. Furthermore, we explore whether the generated uncertainty maps can be used for more fine-grained quality estimation of the generated sCT images by investigating the correlation between the uncertainty and the quality of the produced sCT.

\begin{figure}[t]
    \centering
    \includegraphics[width=0.85\textwidth]{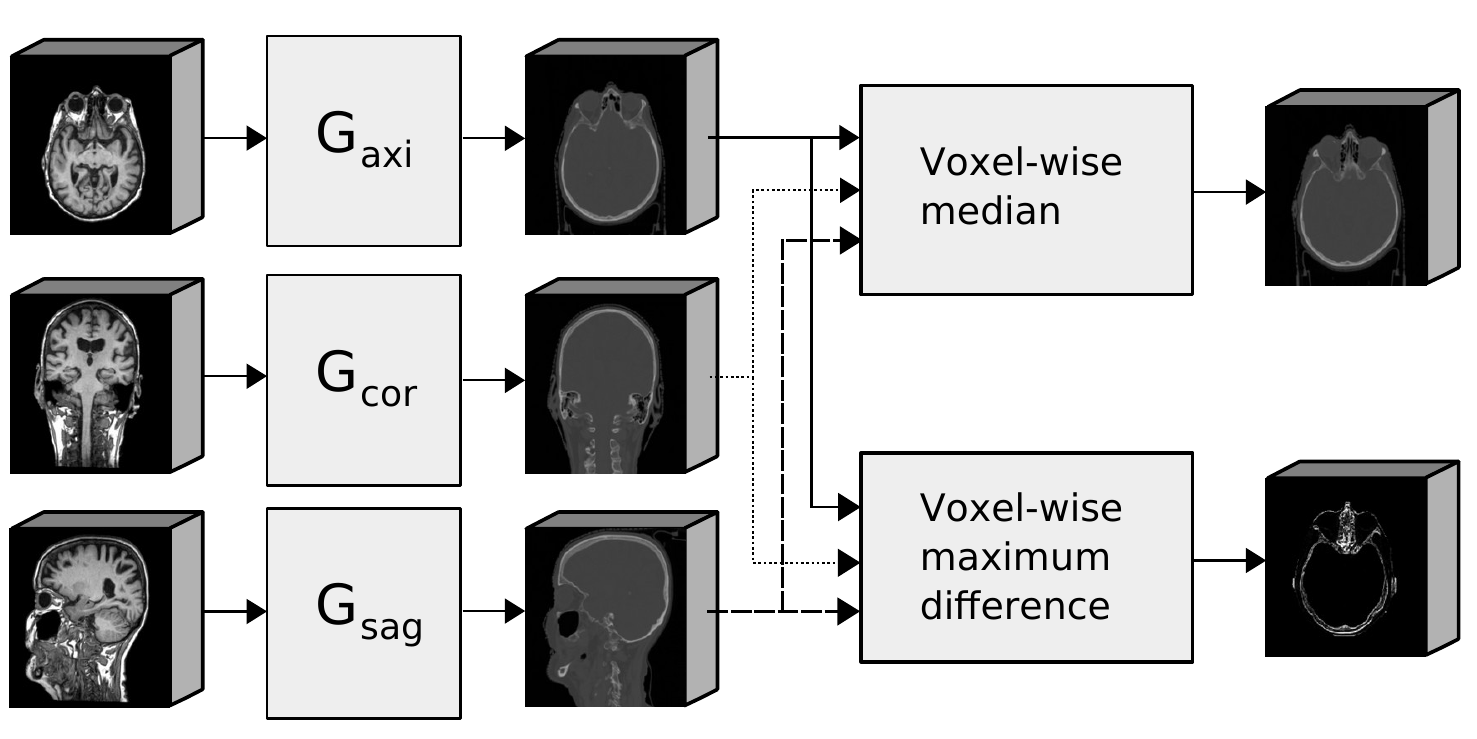}
    \vspace{0.35cm}
    \caption{A schematic overview of the proposed processing pipeline. G\textsubscript{axi}, G\textsubscript{cor} and G\textsubscript{sag} are CT generative networks trained on axial, coronal and sagittal slices, respectively. The voxel-wise median among the three predictions is taken as the sCT result, while the maximum voxel-wise difference yields an uncertainty map.}
    \label{fig:test_pipeline}
\end{figure}

\section{Data}\label{sec:data}
This study includes brain MR and CT images of 52 patients who were scanned at the University Medical Center Utrecht (Utrecht, the Netherlands) for RT treatment planning. For each patient, planning T\textsubscript{1}-weighted MR brain volumes were available, both with and without gadolinium-based contrast agent, acquired using a Philips Ingenia 1.5T MR system. Volumes were acquired with a voxel size of $1.1\times1.1\times1.0$ mm$^3$, 8\textdegree~ flip angle, 7 ms repetition time, and 3.1 ms echo time. Scans were reconstructed to a voxel size of $0.9\times0.9\times1.0$ mm$^3$. For each MR volume, a matching CT volume was available, helically acquired on a Philips Brilliance Big Bore scanner at 120 kVp, 450 mA. The CT volumes were reconstructed with a slice thickness of 1 mm and in-plane resolutions varying between 0.7 and 1.0 mm$^2$. The MR volumes in this set were resampled and rigidly registered to the CT volumes. In the rest of this work, we refer to the MR volumes without gadolinium-based contrast agent from this set as the \textit{RT} set, and to the MR volumes with contrast agent as the \textit{RT}\textsubscript{\textit{gado}} set.

In addition, we included 34 3T T\textsubscript{1}-weighted MR brain volumes of healthy volunteers included in the OASIS project \cite{marcus2007open}. These volumes were acquired on a Siemens scanner with a voxel size of $1.0\times1.0\times1.25$ mm$^3$ and were vertically resampled to an isotropic voxel size of 1.0 mm. No CT volumes were available for these MR volumes. We refer to this set as the \textit{OASIS} set.

\begin{figure}[t]
    \centering
    \includegraphics[width=0.6\textwidth, clip, trim=0mm 0mm 0mm 0mm]{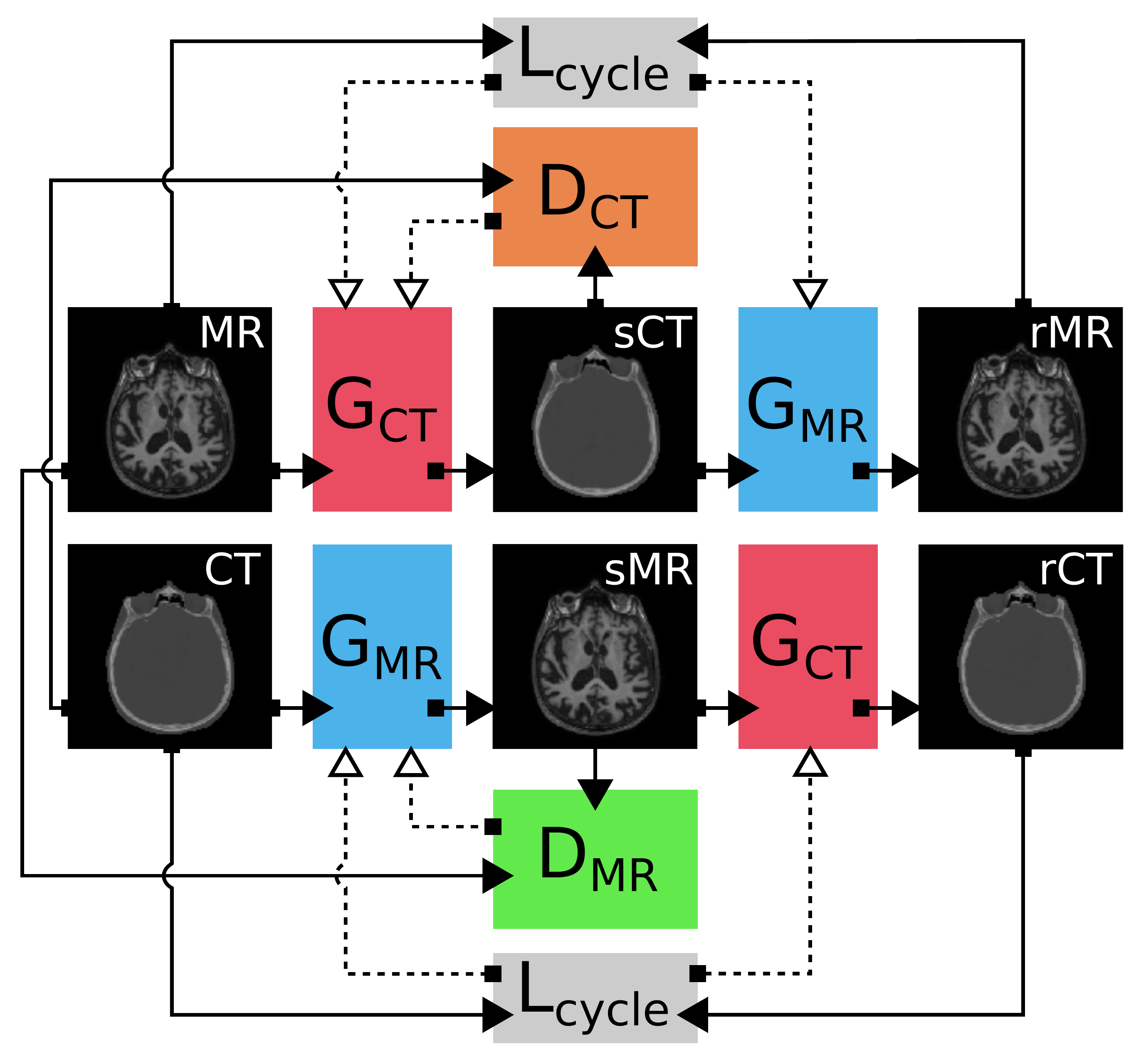}
    \vspace{0.35cm}
    \caption{A schematic overview of the CycleGAN configuration. Solid arrows indicate the flow of data, dashed lines indicate the loss terms for the generators. The generators ($G_{CT}$ and $G_{MR}$) are used to create a synthetic CT and a synthetic MR respectively, shown in the middle of the figure. The opposite generators (duplicated in the figure for visual simplicity) are used to recreate the original input images (yielding rMR and rCT). Two discriminators ($D_{CT}$ and $D_{MR}$) are trained simultaneously to differentiate between real and synthetic images; the generators are trained to fool these discriminators via adversarial loss terms. The cycle-consistency loss terms (L1 norms of MR-rMR and CT-rCT) encourage the reconstructed images to match with the original images. }
    \label{fig:cyclegan_schematic}
\end{figure}

\begin{figure}[t]
    \centering
    \includegraphics[width=0.95\textwidth, clip, trim=30mm 20mm 15mm 16mm]{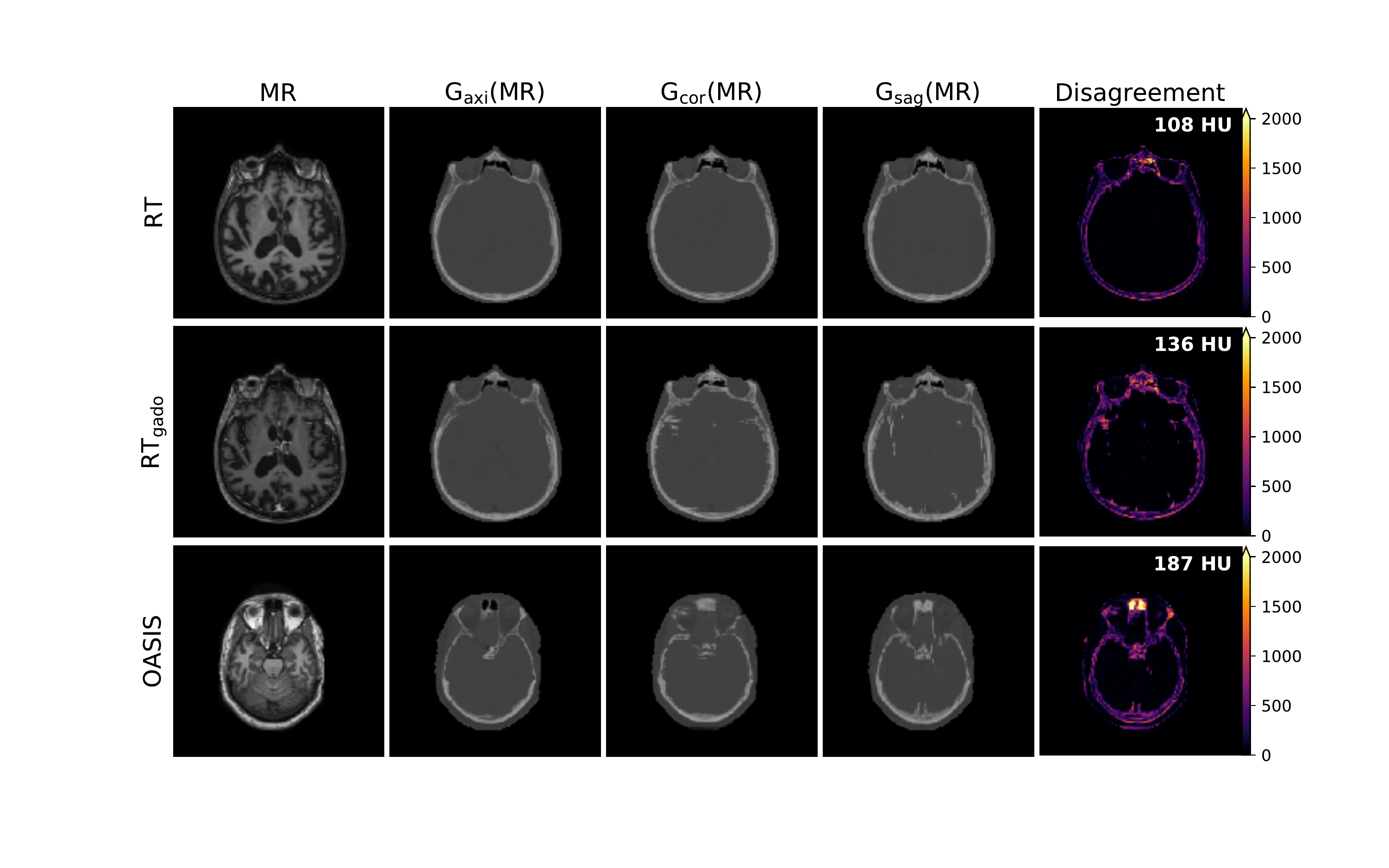}
    \vspace{0.35cm}
    \caption{Results from each of the three test sets. From left to right: input image, results generated using G\textsubscript{axi}, G\textsubscript{cor} and G\textsubscript{sag}, and the resulting uncertainty map. Inputs from top to bottom: \textit{RT} MR (in-distribution), \textit{RT}\textsubscript{\textit{gado}} MR (out-of-distribution) and \textit{OASIS} MR (out-of-distribution). The upper two rows show matched slices from the same patient. Values in the upper right corner indicate the mean uncertainty within the body contour for the shown slices. Note that the models were only trained using images in the \textit{RT} data set.}
    \label{fig:result_images}
\end{figure}

\section{Methods}\label{sec:methods}
We trained an ensemble of three MR-to-CT neural networks: G\textsubscript{axi}, G\textsubscript{cor} and G\textsubscript{sag}, which generate sCT slices in the axial, coronal and sagittal plane, respectively. These networks were trained using the CycleGAN configuration proposed by Zhu et al. \cite{zhu2017unpaired} to learn a mapping between the input MR slices and CT slices as per Wolterink et al.\cite{wolterink_deep_2017}. A schematic overview of the CycleGAN configuration is shown in Fig.~\ref{fig:cyclegan_schematic}. The configuration contains two generator networks that are trained to learn a directional mapping between the MR and CT domains, along with two discriminator networks trained to distinguish real images from each domain from the images produced by the generator networks. At the same time, the generators are trained to fool the discriminators, with the aim of synthesizing realistic generated images. A cycle-consistency loss term is added to the generator loss, which maximizes the amount of information preserved when mapping a generated image back to its original domain. This encourages the generators to map the information from the input domain onto the output domain in a way that results in realistic images. Once the networks are trained, the two generators can be used separately to map images from one domain to the other. In our experiments, only the MR-to-CT generators of the trained CycleGAN models are used. Note that three separate instances of this configuration were trained: one for each anatomical plane.

At test time, an input MR image is processed using G\textsubscript{axi}, G\textsubscript{cor}, and G\textsubscript{sag}, resulting in three 3D sCT images. These sCT images are combined into one result by taking the median Hounsfield unit (HU) at each voxel. Moreover, for each voxel the uncertainty is determined as the maximum absolute difference of the intensity values between any two of the three network results (Fig.~\ref{fig:test_pipeline}). To obtain one metric for uncertainty in a generated sCT, we average the uncertainty for all voxels inside of a body contour. This leads to an uncertainty metric that is insensitive to the volume of the head. The body contours are automatically extracted from the input MR volumes using thresholding and morphological region filling.

\section{Experiments and Results}
The \textit{RT} set was divided into a training set of 30 patients, a test set of 20 patients and a validation set of 2 patients. Using the training set, three CycleGAN models were trained using the training scheme proposed in Zhu et al.\cite{zhu2017unpaired} Each model was trained using 2D slices from one imaging plane, to learn a mapping between the MR volumes without contrast agent and the CT volumes. The models were trained using randomly cropped patches of at most 256\textsuperscript{2} pixels. During inference, slices were constructed by applying the network to patches from each of the corners of the slice and stitching the resulting sCT patches together. In regions where multiple of these corner-patches overlapped, the HU values from the first patch to finish processing were selected.

The MR-to-CT generative models were used to compute sCT images and uncertainty maps from the three test sets: 20 RT MR volumes without contrast agent (\textit{RT}), 20 RT MRs with gadolinium contrast agent (\textit{RT}\textsubscript{\textit{gado}}) and 34 OASIS MR volumes (\textit{OASIS}). As the models were only trained using images from the distribution of the \textit{RT} set, the last two sets contain out-of-distribution MR inputs to the generative models. An example slice from each test set is shown in Fig.~\ref{fig:result_images}, along with the corresponding slices from the sCT images produced by the three generators and the uncertainty map resulting from these slices. 

\begin{figure}[t]
    \centering
    \includegraphics[height=6cm, clip, trim=4mm 3.5mm 0 5mm]{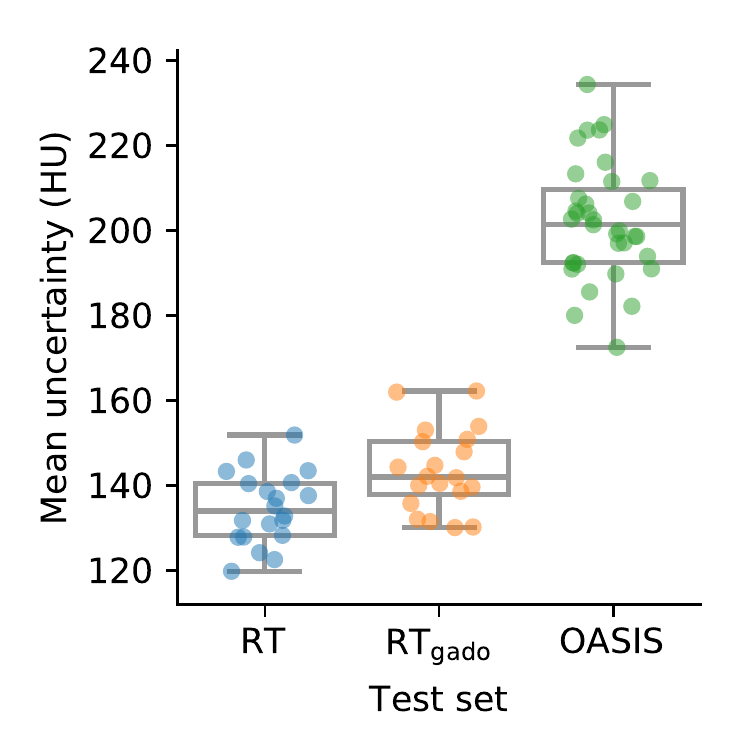}
    \vspace{0.35cm}
    \caption{Mean ensemble uncertainty for each patient in all three test sets. Each data point corresponds to the uncertainty in a single test image, averaged over all voxels within the body contour. }\label{fig:disagreement_setwise}
\end{figure}

We compute the average uncertainty within an automatically extracted body contour. This yields a single uncertainty value for each generated sCT. The average uncertainties for all images in the three data sets are shown in Fig.~\ref{fig:disagreement_setwise}. An unpaired t-test shows that the average uncertainty in the \textit{RT}\textsubscript{\textit{gado}} (144\textpm 9.6 HU) and \textit{OASIS} (202\textpm13.7 HU) sets is significantly (p \textless~0.05) higher than uncertainty in the \textit{RT} set (135\textpm 8.3 HU). Note that based only on the average uncertainty in the result image, images in the out-of-distribution OASIS set could be identified with 100\% accuracy.

\begin{figure}[t]
    \centering
    \includegraphics[height=6cm, clip, trim=0 3.5mm 0 3.5mm]{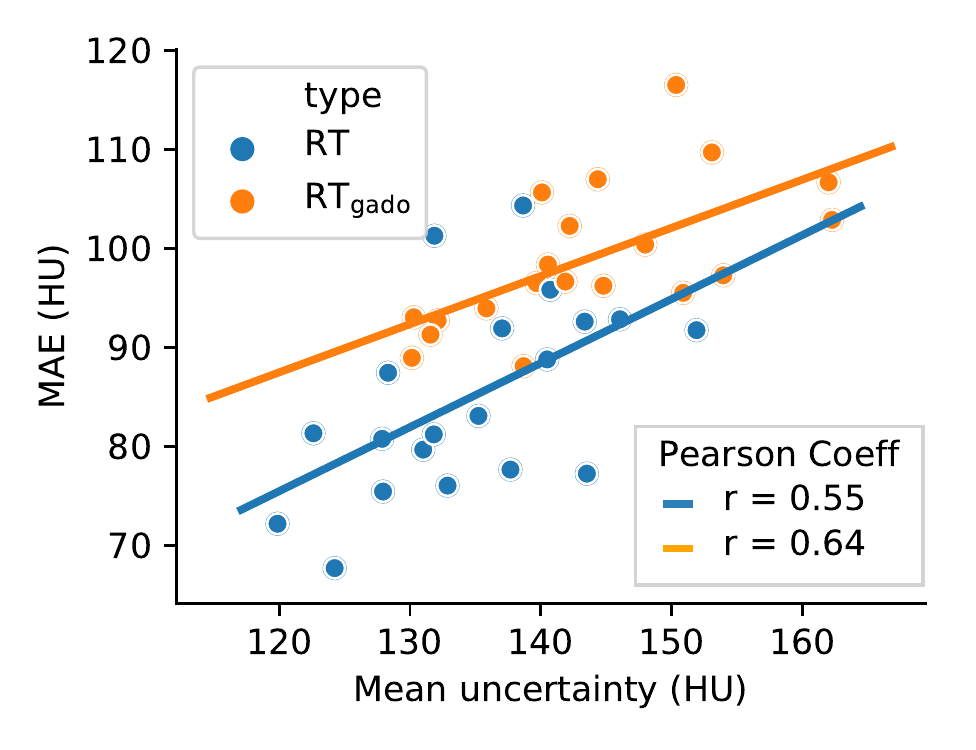}
    \vspace{0.35cm}
    \caption{Correlation between the mean ensemble uncertainty and the mean absolute error in the \textit{RT} and \textit{RT}\textsubscript{\textit{gado}} data sets. Lines indicate linear correlation estimates for both sets. The corresponding Pearson coefficients are indicated in the bottom right.}\label{fig:mae_correlation}
\end{figure}

The results in Fig. \ref{fig:disagreement_setwise} indicate that uncertainty is higher for out-of-distribution images, but this does not mean that the resulting images are necessarily wrong. Therefore, we performed an additional experiment in which we assessed the correlation between uncertainty and errors in the synthetic CT image. We calculated the mean absolute error (MAE) between sCT and reference CT images in the \textit{RT} and \textit{RT}\textsubscript{\textit{gado}} sets; the results are shown in Fig.~\ref{fig:mae_correlation}. These results confirm that there is a clear correlation between the average ensemble uncertainty and the MAE in both sets for which reference CT images are available. Additionally, as this correlation holds for images within the \textit{RT} set, these results indicate that ensemble uncertainty could also be valuable for quality control when the input is from the correct distribution. Furthermore, this figure shows there is some overlap between the MAE distributions of the \textit{RT} and the \textit{RT}\textsubscript{\textit{gado}} sets, indicating that the trained models are somewhat robust against the presence of gadolinium in the input images.

\section{Discussion}
In our experiments we explored the use of an ensemble of sCT generators as a method for automatic online quality control in MR-only radiotherapy workflows. By considering the disagreement of different sCT generators as a metric for their collective uncertainty, the networks can automatically alert a human operator if the uncertainty becomes higher than a threshold. The obvious fault mode of such a system is an input for which all networks in the ensemble make the same mistake. In this work, such a fault mode was mitigated by training the networks on different views on the data (i.e. axial, coronal and sagittal slices), causing the generators to learn mappings between different input and output distributions. This should reduce the chance that an incorrect mapping by one of the networks coincides with identical incorrect mappings by the other two. Our experiments have shown that there is indeed a negative correlation between the disagreement between the generators and the sCT quality.

A curious observation from Fig.~\ref{fig:mae_correlation} is that the distributions of mean absolute errors for the \textit{RT} set the \textit{RT}\textsubscript{\textit{gado}} set have substantial overlap. This indicates that our sCT generative models are somewhat robust to the presence of gadolinium-based contrast agent in the MR images, even though no images with contrast agent were present in the training set. This can explain the overlap of the distributions of mean uncertainties in these sets as well.
Conversely, when comparing the uncertainty results of the \textit{RT} set and the \textit{OASIS} set in Fig.~\ref{fig:disagreement_setwise}, we observe that these sets are perfectly separable on the mean uncertainty metric. This means that all images from the set of incompatible MR images could be caught as invalid inputs to the sCT generative method using a simple threshold.

While the presented single value for uncertainty has merit in practical validation systems due to its simple nature, the absence of spatial information can be considered a weakness of the method. In the radiotherapy treatment workflow, the accuracy of an sCT is more relevant in the sections that are irradiated during treatment. This implies it could be valuable to acquire individual quality estimates for different regions, as errors in irrelevant sections of the sCT could be acceptable. Future work could investigate the use of spatial uncertainty information to find such a localised estimate of the sCT quality.

\section{Conclusion}
In this work, we have developed a method to aid quality control of MR-to-CT generative deep-learning models, to facilitate their clinical adoption in radiotherapy workflows. We have shown the automatically generated uncertainties produced by our method correlate with the quality of the generated synthetic CTs, and that they could be used to detect input images that are not from the correct data distribution. Such an automated check would be highly valuable as an automated online validation step in the clinical workflow.

\section{New or breakthrough work to be presented}
We have presented a method for estimating the uncertainty of an MR-to-CT generative deep-learning model, as used in MR-only radiotherapy workflows. We have shown that our estimated uncertainties correlate well with the quality of the generated sCTs, paving the way for automatic quality control in clinical practice.

\bibliography{spielib_qa} %
\bibliographystyle{spiebib} 

\end{document}